\documentclass[reprint,aps,prapplied,superscriptaddress,twocolumn]{revtex4-2}
\usepackage{soul}
\usepackage{soul}
\usepackage{soul}
\usepackage{comment}
\usepackage{amsmath}
\usepackage{graphicx}
\usepackage{color}
\usepackage{adjustbox}
\usepackage{hyperref}
\usepackage{multirow}
\usepackage{cancel}
\usepackage{pdfpages}
\usepackage{etoolbox}

\makeatletter
\patchcmd{\@outputpage@head}{\@ifx{\LS@rot\@undefined}{}{\LS@rot}}{}{}{}
\makeatother

\begin{document}

%\preprint{}-

\title{Quantum sensing of \textcolor{black}{time-dependent} magnetic signals with molecular spins}

% \email, \thanks, \homepage, \affiliation all apply to the current
% author. Explanatory text should go in the []'\s, actual e-mail

% other informationI
%\email[]{}
\author{Matteo Lanza}
\email[]{matteo.lanza@unimore.it}
\affiliation{Dipartimento di Scienze Fisiche, Informatiche e Matematiche, Università degli Studi di Modena e Reggio Emilia, via G. Campi 213/A, 41125 Modena, Italy}
\affiliation{Istituto Nanoscienze CNR, Centro S3, via G. Campi 213/A, 41125 Modena, Italy}
\author{Claudio Bonizzoni}
\affiliation{Dipartimento di Scienze Fisiche, Informatiche e Matematiche, Università degli Studi di Modena e Reggio Emilia, via G. Campi 213/A, 41125 Modena, Italy}
\affiliation{Istituto Nanoscienze CNR, Centro S3, via G. Campi 213/A, 41125 Modena, Italy}
\author{Olga Mironova}
\affiliation{Dipartimento di Scienze Chimiche e Geologiche e UdR INSTM, Università degli Studi di Modena e Reggio
Emilia, via G. Campi 103, 41125 Modena, Italy}
\author{Fabio Santanni}
\affiliation{Dipartimento di Chimica "U. Schiff", Università di Firenze, Via della Lastruccia 3, Sesto F.no, 50019, Italy}
\author{Alessio Nicolini}
\affiliation{Dipartimento di Scienze Chimiche e Geologiche e UdR INSTM, Università degli Studi di Modena e Reggio
Emilia, via G. Campi 103, 41125 Modena, Italy}
\author{Alberto Ghirri}
\affiliation{Istituto Nanoscienze CNR, Centro S3, via G. Campi 213/A, 41125 Modena, Italy}
\author{Andrea Cornia}
\affiliation{Dipartimento di Scienze Chimiche e Geologiche e UdR INSTM, Università degli Studi di Modena e Reggio
Emilia, via G. Campi 103, 41125 Modena, Italy}
\author{Marco Affronte}
\affiliation{Dipartimento di Scienze Fisiche, Informatiche e Matematiche, Università degli Studi di Modena e Reggio Emilia, via G. Campi 213/A, 41125 Modena, Italy}
\affiliation{Istituto Nanoscienze CNR, Centro S3, via G. Campi 213/A, 41125 Modena, Italy}
%\thanks{}
%\affiliation{}

\date{\today}

\begin{abstract}   
Molecular spins offer a promising platform for quantum sensing, particularly in organic, supramolecular or biological environments. 
Recognition of the signals by these systems is of particular interest given their possible integration into more complex structures and their possible use as sensors in close proximity to analytes. In this work, we develop two quantum sensing protocols that enable discrimination between different time-dependent magnetic field, without requiring its periodicity to match with the microwave manipulating sequence. These are based on the Hahn echo sequence and have been tested on VO(TPP) and VOPt(SOCPh)\textsubscript{4} molecular spins embedded in a superconducting YBCO microwave planar resonator. We report a magnetic field sensitivity up to 2.57 $\cdot 10^{-7}$ T Hz\textsuperscript{$-\frac{1}{2}$} (with lower bounds approaching 2.87 $\cdot 10^{-8}$ T Hz\textsuperscript{$-\frac{1}{2}$}) for signals with duration of a few  microseconds. Under the given conditions, the minimum signal area that can be measured is in the $10^{-10}$ T s range, suggesting a potential trade-off between minimum measurable field and the required signal duration and memory time. 
\end{abstract}

%\keywords{}

\maketitle

%%%%%%%%%%%%%%%%%%%%%%%%%%%%%%%%%%%%%%%%%%%%%%%%%%%%%%%%%%%%%%%%%%%%%%%%%%%%%%%%%%%%%%%%%%%%%%%%%%%%%
\section{Introduction}
Quantum sensing exploits the quantum properties of a system for detecting a physical quantity.  Nuclear and electron spin systems with sufficiently long coherence times have the potential to be used for this purpose \cite{degenREVMODPHYS2017}.
Nitrogen-Vacancy (NV) centers in diamond are nowadays prototypical examples having shown remarkable performances \cite{bar-gillSolidstateElectronicSpin2013,herbschlebUltralongCoherenceTimes2019, zhouPRX2020, schirhaglANNREVPHYSCHEM2014}  including single spin detection through Optically-Detected-Magnetic-Resonance (ODMR) \cite{taylorNATPHYS2008, zhouPRX2020} and microwave (MW) Dynamical Decoupling schemes \cite{phamPRB2012, rondinREPPROGRPHYS2014, taylorNATPHYS2008, fangPRL2013, zhouPRX2020, farfurnikJOURNOFOPT2018}.
A large collection of NV centers acts as an average of the single spin responses \cite{zhouPRX2020} and gives sufficiently large luminescent signals. This allows to reach typical sensitivities of the order of few $10^{-9}$ T Hz\textsuperscript{$-\frac{1}{2}$} when detecting AC magnetic fields whose period is matched with the free precession time of the protocols \textcolor{black}{with dynamical decoupling sequences \cite{fangPRL2013,phamPRB2012,farfurnikJOURNOFOPT2018} and down to $10^{-12}$ T Hz\textsuperscript{$-\frac{1}{2}$} with the use of a continuous heterodyne detection scheme \cite{WangScience2022}.} \\
Molecular spins, in addition to displaying microsecond coherence times \cite{atzorimorraJACS2016}, greatly benefit from chemical design, a powerful way to control their composition, their local environment, tune their energy levels, incorporate them in supramolecular structures or crystalline 3D or 2D lattices. \cite{timcoEngineeringCouplingMolecular2009, santanniMetalloporphyrinsBuildingBlocks2024, yamabayashiJACS2018}.  
Their portability, i.e. the capability of the sensor to be attached to an analyte with atomic precision at nm-distance, is one of their most interesting features. In fact, molecular spin labels have already been used to measure distances within a single protein, in combination with Pulsed Electron-Electron Double Resonance (PELDOR) sequences \cite{Jeschke2006}.
In this respect, they constitute a valid \textcolor{black}{ resource} for spin-based quantum sensing \cite{yuMolecularApproachQuantum2021, latawiecDetectingChiralityinducedSpin2025, dzubaPulsedEPRMethod2007}
and the development of sensing protocols can expand their potential in this field \cite{troianiJMMM2019, sunJACS2022, singhPhysRevResearch2025, singh2024highsensitivitypressuretemperature, priviteraRoomTemperatureOpticalSpin2025a, grzegorzekMetalatedPorphyrinStable2020, bonizzoniQuantumSensingMagnetic2024}. 
We have shown the possibility of using molecular spins as quantum memories \cite{bonizzoniNPJQUANT2020} and, more recently, for quantum sensing of AC magnetic fields \cite{bonizzoniQuantumSensingMagnetic2024}. Along this line, the detection of a non-periodic time-dependent magnetic field signal would allow the recognition of specific events. However, due to the higher number of harmonics expected and to the fact that the microwave protocols cannot match their period, this is a more complex task. For instance, this has been demonstrated in state-of-the-art experiments on NV centers in diamond through ODMR protocols \cite{cooperNATCOMM2014, zopesReconstructionfreeQuantumSensing2019a, cooperTimeresolvedMagneticSensing2014, magesanReconstructingProfileTimevarying2013}. Yet, the sequences used in these works employ optical initialization and detection \cite{mazeNanoscaleMagneticSensing2008}, several MW pulses \cite{magesanReconstructingProfileTimevarying2013}, multiple triggering of the external signal \cite{zopesReconstructionfreeQuantumSensing2019a} or a combination of these elements.\\
In this work, we propose two spin echo sequences that consist of only two MW pulses and do not require optical readout. Furthermore, the external magnetic signal has to be triggered only once per sequence repetition \textcolor{black}{and it does not need to match the total sequence time, nor a specific time alignment with the MW sequence}. For both protocols, based on a Hahn echo sequence {\cite{hahnPHYSREV1950}}, the microwave pulses or the external signal are moved to allow for the discrimination of different signal shapes. The sensing principle relies on the difference in the phase accumulation obtained during the two free precession times of the sequence at successive steps. We perform our tests on molecular spins at liquid Helium temperature using planar superconducting \textcolor{black}{MW} resonators. However, we notice that our protocols can also be implemented up to room temperature and in a commercial spin resonance spectrometer, with potential application to a variety of spin systems.

%%%%%%%%%%%%%%%%%%%%%%%%%%%%%%%%%%%%%%%%%%%%%%%%%%%%%%%%%%%%%%%%%%%%%%%%%%%%%%%%%%%%%%%%%%%%%%%%%%%%%
\section{Experimental details} \label{sec:Exp}
\begin{figure*}
\includegraphics[width=\textwidth]{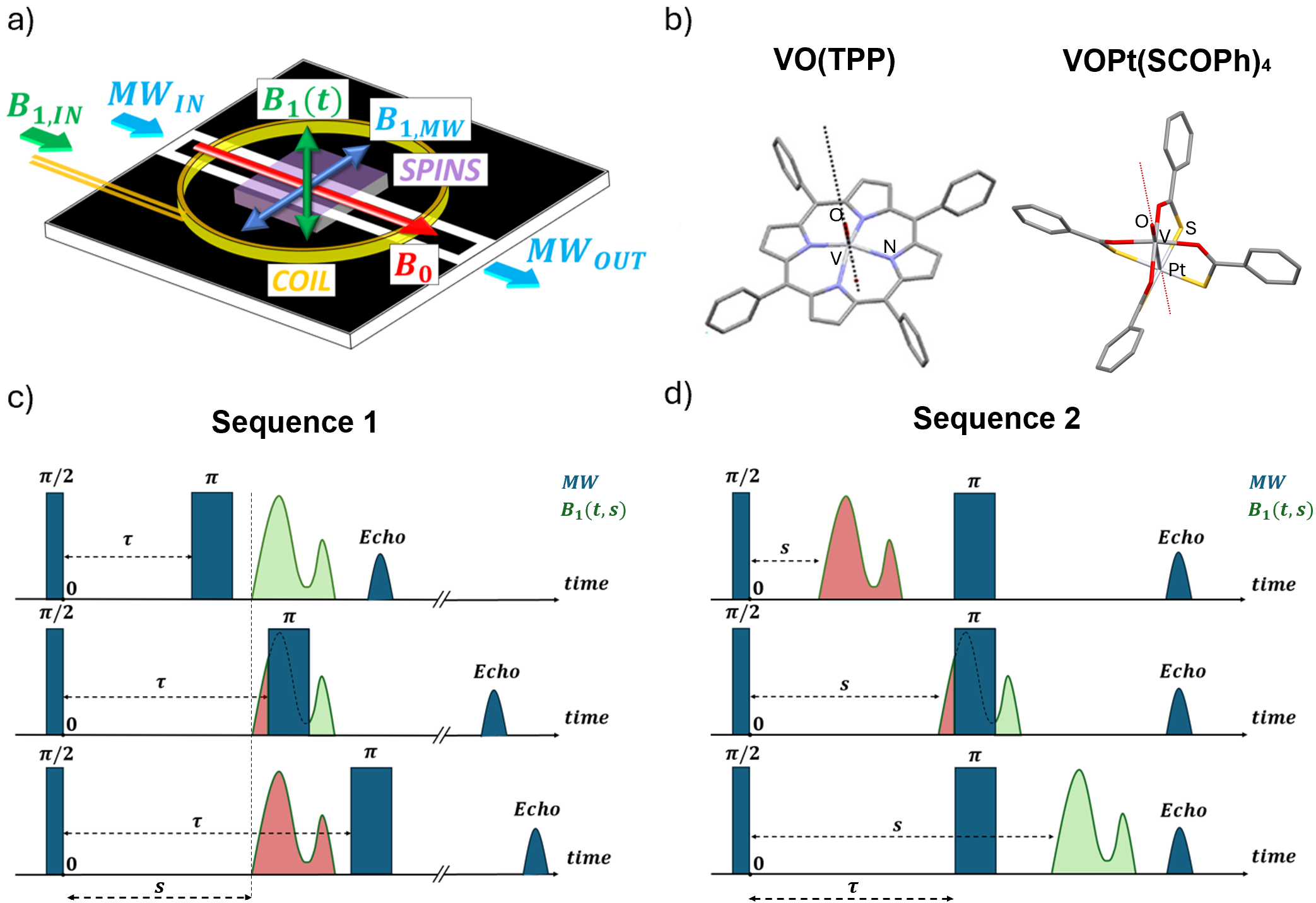}%
\caption{(a) YBCO superconducting coplanar resonator used to deliver the MW pulses, $B$\textsubscript{1,MW}. The sample is placed in the middle of the resonator and it is surrounded by a RF copper coil used to deliver the external magnetic field to be measured, $B$\textsubscript1($t$). The static field, $B_0$, is applied along the longitudinal axis of the resonator. Figure reproduced from \cite{bonizzoniQuantumSensingMagnetic2024} under CC license. (b) Molecular structures of VO(TPP) (left) and VOPt(SOCPh)\textsubscript{4} (right). The picture of the VO(TPP) molecule is adapted from \cite{bonizzoniNPJQUANT2020} under CC license. (c) Sequence 1: the two MW pulses (blue) are used to generate an echo, and the interpulse delay $\tau$ is changed step-by-step while the position of the magnetic field signal, s, is held fixed. (d) Sequence 2: the interpulse delay is kept fixed while the position of the magnetic field signal is swept across the whole sequence by increasing $s$ step-by-step. In both (c) and (d), the phase accumulation is proportional to the difference between the signal areas remaining before and after the $\pi$ pulse (opposite phase precession sign is represented with red and green shaded areas).\label{fig:FIG1}}
\end{figure*}

\subsection{Setup}
A high-$T$\textsubscript{C} YBCO superconducting coplanar resonator is used for the manipulation and readout of the spin system with MW pulse sequences. The resonator has a bandwidth of a few MHz (\textcolor{black}{2}-10 MHz) throughout the experiments. This value can be tuned at room temperature by changing the distance between the MW feed antenna and the resonator. \textcolor{black}{Possible effects/limitations related to the resonator bandwidth on the pickup of the echo signals are discussed in the Supplementary Information \cite{supp}.} The sample is placed at the center of the resonator, where the generated MW magnetic field, $B$\textsubscript{1,MW}, has its maximum\textcolor{black}{, as in Fig. \ref{fig:FIG1} (a)}. In addition, a copper coil is placed on top of the resonator, \textcolor{black}{surrounding }the sample, to generate \textcolor{black}{a calibrated} magnetic signal, $B$\textsubscript{1}($t$). A static magnetic field, $B$\textsubscript{0}, is applied along the \textcolor{black}{longitudinal} axis of the resonator. In this configuration, all three fields ($B$\textsubscript{0}, $B$\textsubscript{1,MW} and $B$\textsubscript{1}($t$) are perpendicular to each other, Fig. \ref{fig:FIG1}) \cite{bonizzoniQuantumSensingMagnetic2024}. The MW pulses are generated using a homemade heterodyne setup in which a mixer modulates a \textcolor{black}{GHz} frequency ($\nu$\textsubscript{0} - $\nu$\textsubscript{IF}) with a lower $\nu$\textsubscript{IF} = 70 MHz frequency carrying the desired pulse pattern \cite{bonizzoniNPJQUANT2020}. The readout of the system is performed through a spin echo, which can be detected in both phase and amplitude by using a second mixer \textcolor{black}{receiving} $\nu$\textsubscript{0} - $\nu$\textsubscript{IF} as a reference frequency. Both $\nu$\textsubscript{IF} and the signal for the coil are generated using two different channels of an Arbitrary Waveform Generator (AWG) \cite{bonizzoniQuantumSensingMagnetic2024}. A Quantum Design Physical Property Measurement System (QD PPMS) is used for cooling the sample to 2-3.5 K and applying the static magnetic field in the plane of the YBCO resonator.

\subsection{Sequences}

%%%%%%%%%%%%%%%%%%%%%%%%
\begin{figure*}
\includegraphics[width=\textwidth]{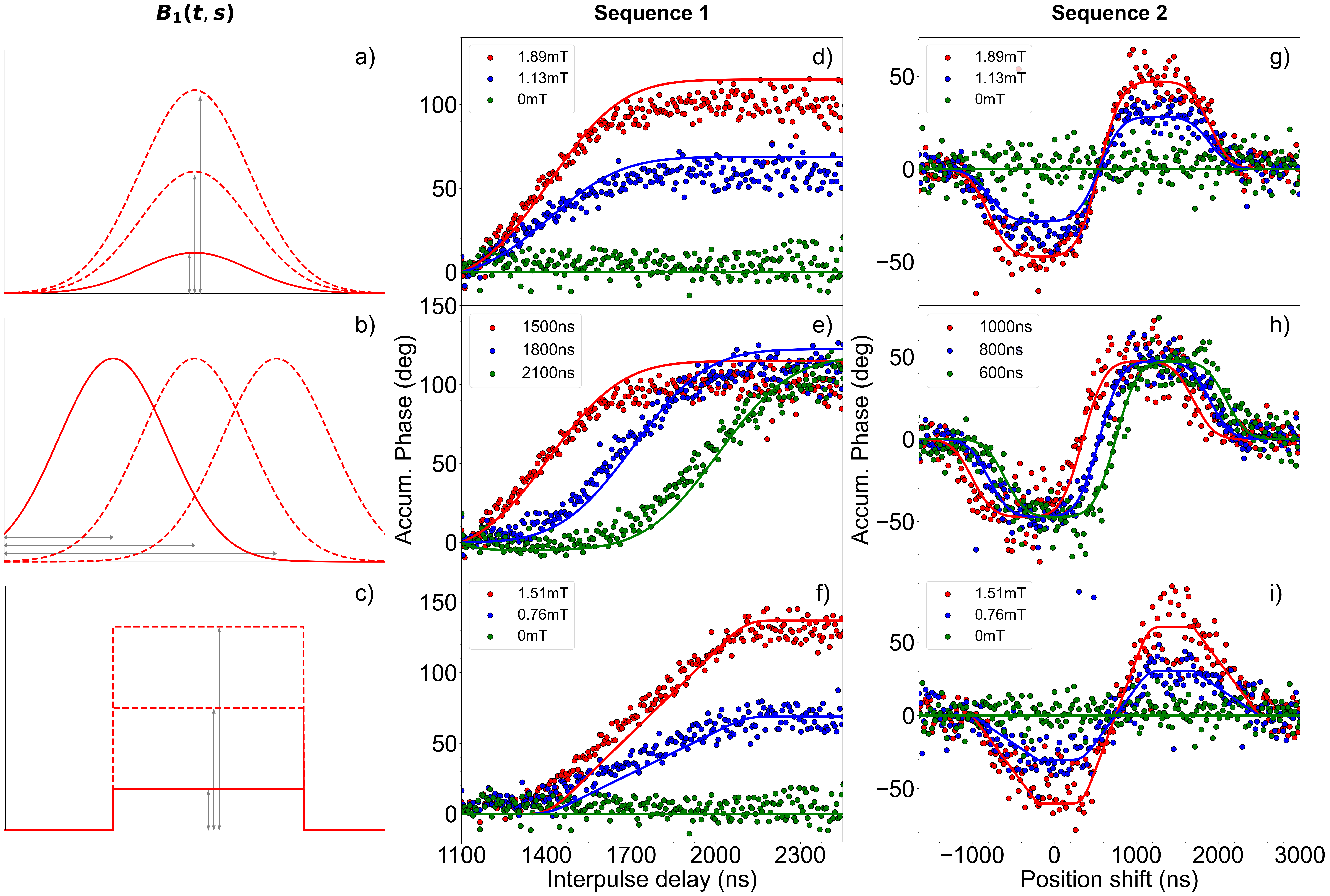}

\caption{Sequences 1 and 2 performed on VO(TPP) at 2-3.5 K and at a fixed value of the static field $B$\textsubscript0. The protocols are tested with a Gaussian signal (a,b) or with a rectangular signal (c). (d,g) Phase accumulation measured for different amplitudes, $B$\textsubscript{1,max}, of the Gaussian signal for Sequence 1 (d) and 2 (g). (e,h) Phase accumulation measured when the central position, $t$\textsubscript0, of the Gaussian signal is changed in both sequence 1 (e) and 2 (h). (f, i) The amplitude, $B$\textsubscript{1,max}, of the rectangular signal is changed in both Sequence 1 (f) and 2 (i). \textcolor{black}{For Sequence 1 an experimental baseline has been removed (see Supplementary Information for details \cite{supp}).} In all panels experimental data are shown with dots, while solid lines represent calculated curves obtained with Eq. \ref{eq:phase2}.
\label{fig:FIG2}}
\end{figure*}
%%%%%%%%%%%%%%%%%%%%%%%%

The Hahn echo sequence consists of two MW pulses: an initial $\pi/2$ pulse induces a 90° spin rotation on the Bloch sphere; it is followed by a free precession time, $\tau$, and finally by a $\pi$ pulse that causes a refocusing \cite{hahnPHYSREV1950}. When an external magnetic field $B$\textsubscript{1}($t$) is applied \textcolor{black}{using the coil}, the echo undergoes a phase accumulation \cite{degenREVMODPHYS2017}:
\begin{equation}
    \phi\textsubscript{echo}(T\textsubscript{seq}, s) = \int_{0}^{T\textsubscript{seq}} \gamma B\textsubscript{1}(t, s) dt,
    \label{eq:phase}
\end{equation}
where $T$\textsubscript{seq} is the time between the $\frac{\pi}{2}$ pulse and the spin echo (total sequence time) and $\gamma$ is the transduction parameter. Here, we introduce a dependence on $s$, which is a rigid time shift of $B$\textsubscript{1}($t$) with respect of the end of the $\pi$/2 pulse and this is used to move the signal across the MW sequence. 
Eq. \ref{eq:phase} can be split into three different intervals for the phase accumulation: between the $\frac{\pi}{2}$ and the $\pi$ pulse, during the application of the $\pi$ pulse, between the $\pi$ pulse and the spin echo (Fig. \ref{fig:FIG1} (c, d)). In the third region, the accumulated phase has an opposite sign with respect to the first one due to the refocusing introduced by the $\pi$ pulse. In the second region, the phase accumulation is modulated by a sine function due to the rotation of the magnetization on the Bloch sphere \cite{zopesReconstructionfreeQuantumSensing2019a}. 
The expression for the phase accumulation reads as:

\begin{equation}
\begin{split}
    &\phi\textsubscript{echo}(T\textsubscript{seq}, s)\\
    &= \int_{t\textsubscript{$\frac{\pi}{2}$,end}}^{t\textsubscript{$\pi,start$}} \gamma B\textsubscript{1} (t,s) dt \\
    &- \int_{t\textsubscript{$\pi$,start}}^{t\textsubscript{$\pi$,end}} \gamma B\textsubscript{1} (t,s) \sin\left(\frac{t - t\textsubscript{$\pi$,start}- \frac{T\textsubscript{$\pi$}}{2}}{T\textsubscript{$\pi$}/\pi}\right)  dt \\
    &- \int_{t\textsubscript{$\pi$,end}}^{t\textsubscript{echo}} \gamma B\textsubscript{1} (t,s) dt,
\end{split}
\label{eq:phase2}
\end{equation}
 where $t\textsubscript{$\frac{\pi}{2}$,end}$ is the time at which the $\frac{\pi}{2}$ pulse ends, $t\textsubscript{$\pi$,start}$ and $t\textsubscript{$\pi$,end}$ are the start and end times for the $\pi$ pulse, $T$\textsubscript{$\pi$} is the duration of the $\pi$ pulse,  and $t$\textsubscript{echo} is the time at which the echo is observed. Hereafter, we will fix $t$\textsubscript{$\frac{\pi}{2}$,end} = 0 ns.\\
Since the phase accumulated during an Hahn echo sequence depends on the area of the external signal through Eq. \ref{eq:phase2}, it is possible to exploit the opposite sign obtained in the different regions to detect an external non-periodic signal. This can be achieved with two different sequences:
Sequence 1 (Fig. \ref{fig:FIG1} c) consists of a Hahn echo sequence in which  the interpulse delay ($\tau$) is increased step-by-step, shifting the $\pi$ pulse over the external signal $B$\textsubscript{1}($t$, $s$), which remains unchanged in its starting position by fixing $s$. For each step, an echo is recorded and its phase depends on the area of the signal according to Eq. \ref{eq:phase2}. The total duration of the sequence is related to the interpulse delay through the relation: $T$\textsubscript{seq} = 2 $\tau$ + $T$\textsubscript{$\pi$}. 
Conversely, in Sequence 2 (see Fig. \ref{fig:FIG1} (d)), the $\pi$/{2} and $\pi$ pulses have fixed $\tau$ and the external signal is shifted rigidly step-by-step in time by increasing $s$, while the MW sequence remains unchanged. The accumulated phase appearing in Eq. \ref{eq:phase2} now depends on the position shift, $s$, while $T$\textsubscript{seq} remains unchanged.\\
In both Sequence 1 and 2, the key features of the time-dependent $B$\textsubscript{1}($t$,$s$) signal can be derived from the behaviour of the phase of the echo signal through Eq. \ref{eq:phase2}.\\

\subsection{Molecular Spin Samples}

Both samples contain vanadyl complexes ($S$ = $\frac{1}{2}$, $I$ = $\frac{7}{2}$) magnetically diluted in a diamagnetic matrix of their isostructural titanyl analogues. The first sample is a solid solution of VO(TPP) (2 mol\%) in TiO(TPP), used as a powder with an estimated $T$\textsubscript{m} $\approx$ 2 $\mu$s (at 2 K) and a spin density $\rho$ = 2.3 $\cdot 10^{19}$ spin/cm\textsuperscript3 \cite{yamabayashiJACS2018, drewINORGCHIMACTA1984}. The second sample is a solid solution of VOPt(SOCPh)\textsubscript4 \cite{beachHeterotrimetallicLnOVPtComplexes2020,imperatoQuantumSpinCoherence2024} (1 mol\%) in TiOPt(SOCPh)\textsubscript4 $\cdot$ 2THF, used as a single crystal with an estimated $T$\textsubscript{m} $\approx$ 4 $\mu$s (at 3 K) and $\rho$ = 1.0 $\cdot 10^{19}$ spin/cm\textsuperscript3 \cite{belliniThesis}. The two samples give similar Electron Paramagnetic Resonance (EPR) spectra and  the strongest transition line is chosen by fixing the static field value to match the energy of the YBCO resonator. Further information on the samples preparation, orientation and memory time estimation can be found in the Supplementary Information \cite{supp}. 

\section{Results}
We firstly test our sequences by applying a Gaussian-shaped magnetic field signal in the form:
\begin{equation}
    B\textsubscript{1}(t,s) = B\textsubscript{1,max} e^{(-\frac{(t-(t_{0}+s))^{2}}{2\sigma^{2}})},
    \label{eq:Gaussian}
\end{equation}
\noindent
where $B\textsubscript{1,max}=B\textsubscript{1}(t\textsubscript0 + s)$ is the amplitude of the external field at the center of the Gaussian, $t$\textsubscript{0} + $s$ is its central position in time and $\sigma$ is its width. Here, the aim is to distinguish the effects of changing one parameter ($B$\textsubscript{1,max}, $t$\textsubscript{0}, $\sigma$) at a time while the others are kept fixed.\\
As a further test, we use a rectangular-shaped external signal, with equation:
\begin{equation}
    B\textsubscript{1}(t,s) = \begin{cases} 0, & \mbox{if $t$ $<$ $t$\textsubscript{L} + $s$} \\ 
    B\textsubscript{1,max}, & \mbox{if $t$\textsubscript{L} + $s$ $\leq$ $t$ $\leq$ $t$\textsubscript{R} + $s$} \\ 
    0, & \mbox{if $t$ $>$ $t$\textsubscript{R} + $s$} \end{cases}
    \label{eq:Square}
\end{equation}
where $B$\textsubscript{1,max} is the amplitude of the signal and $t$\textsubscript{L} + $s$  and $t$\textsubscript{R} + $s$ are the left and right time extremes with respect to the time origin, respectively.

\subsection{Quantum sensing with VO(TPP)}
A powder sample of magnetically diluted VO(TPP) is used to initially assess the efficiency of the two protocols. Sequences 1 and 2 are repeated with a Gaussian signal (Eq. \ref{eq:Gaussian}) for different values of $B$\textsubscript{1,max} and $t$\textsubscript{0} (Fig. \ref{fig:FIG2} (a, d, g) and (b, e, h)). This allows us to evaluate whether the protocols are capable of sensing changes in the amplitude and position of the external signal. The resulting echo phase signals (Fig. \ref{fig:FIG2} (d, e)) show clear differences in the amount of accumulated phase and in the starting position of the accumulation with respect to timescale. We also test the protocols with a rectangular signal (Eq. \ref{eq:Square}) by varying its amplitude (see Fig. \ref{fig:FIG2} (c, f, i)). The phase accumulation clearly depends on the value of $B$\textsubscript{1,max}.

%%%%%%%%%%%%%%%%%%%%%%%%
\begin{figure}
\includegraphics[width=\columnwidth]{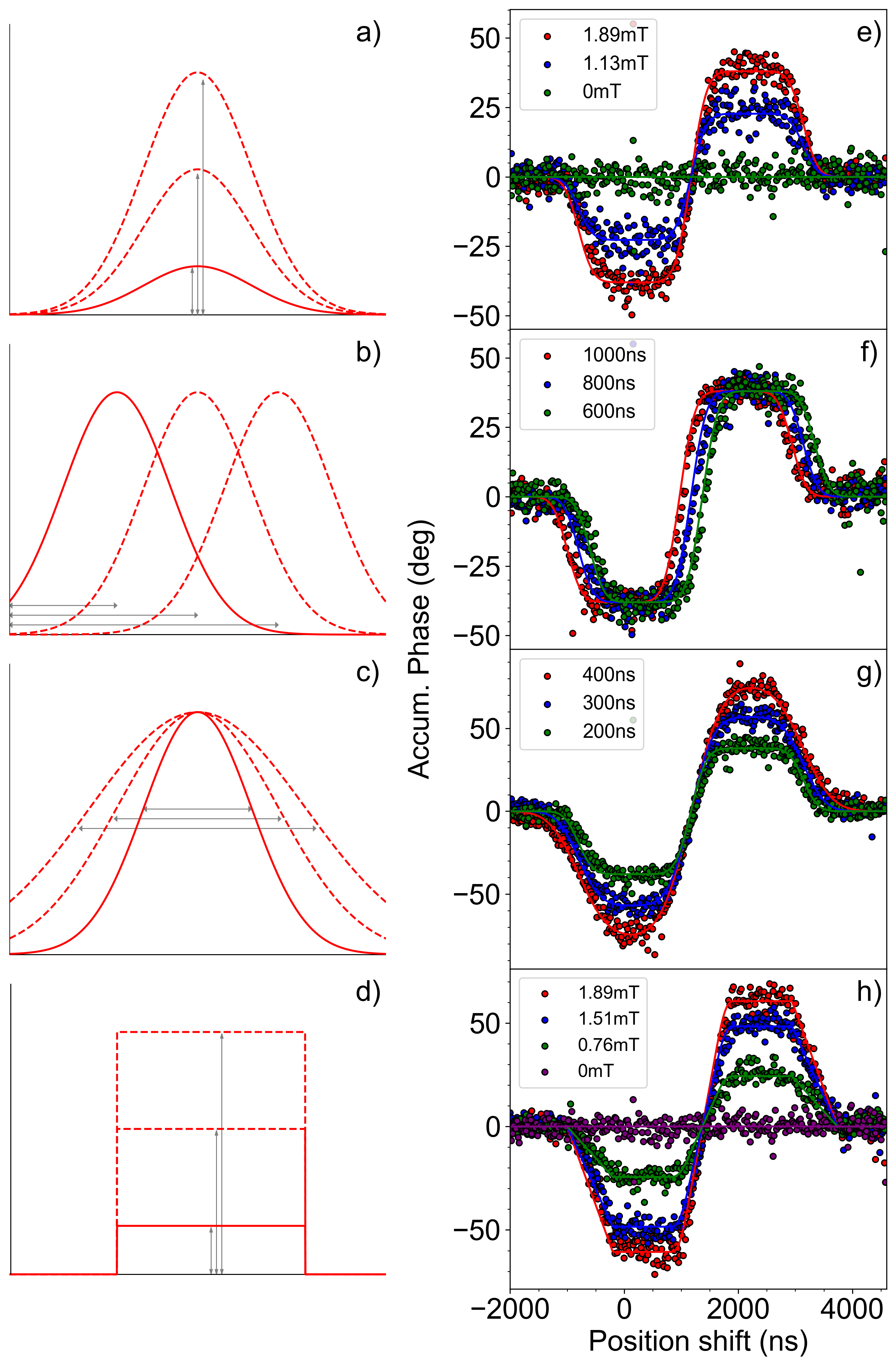}%
\caption{Tests performed with Sequence 2 on VOPt(SOCPh)\textsubscript{4} at $3-3.5K$ and at a fixed static field ($B_0$). A gaussian (a,b,c) and a rectangular (d) signal are used. Phase accumulation (e,f,g) measured when the amplitude $B$\textsubscript{1,max} (e), the central position $t$\textsubscript0 (f) and the width $\sigma$ of the Gaussian signal (g) are progressively increased. (h) Phase of the echo measured for different amplitudes $B$\textsubscript{1,max} of a rectangular pulse signal. In all panels, the dotted traces represent experimental points, while the solid lines represent calculated curves obtained according to Eq. \ref{eq:phase2}).}
\label{fig:FIG3}
\end{figure}
%%%%%%%%%%%%%%%%%%%%%%%%

%\begin{multicols}
%\end{multicols}
\subsection{Quantum sensing with VOPt(SOCPh)\textsubscript{4}}
\begin{figure}
\includegraphics[width=\columnwidth]{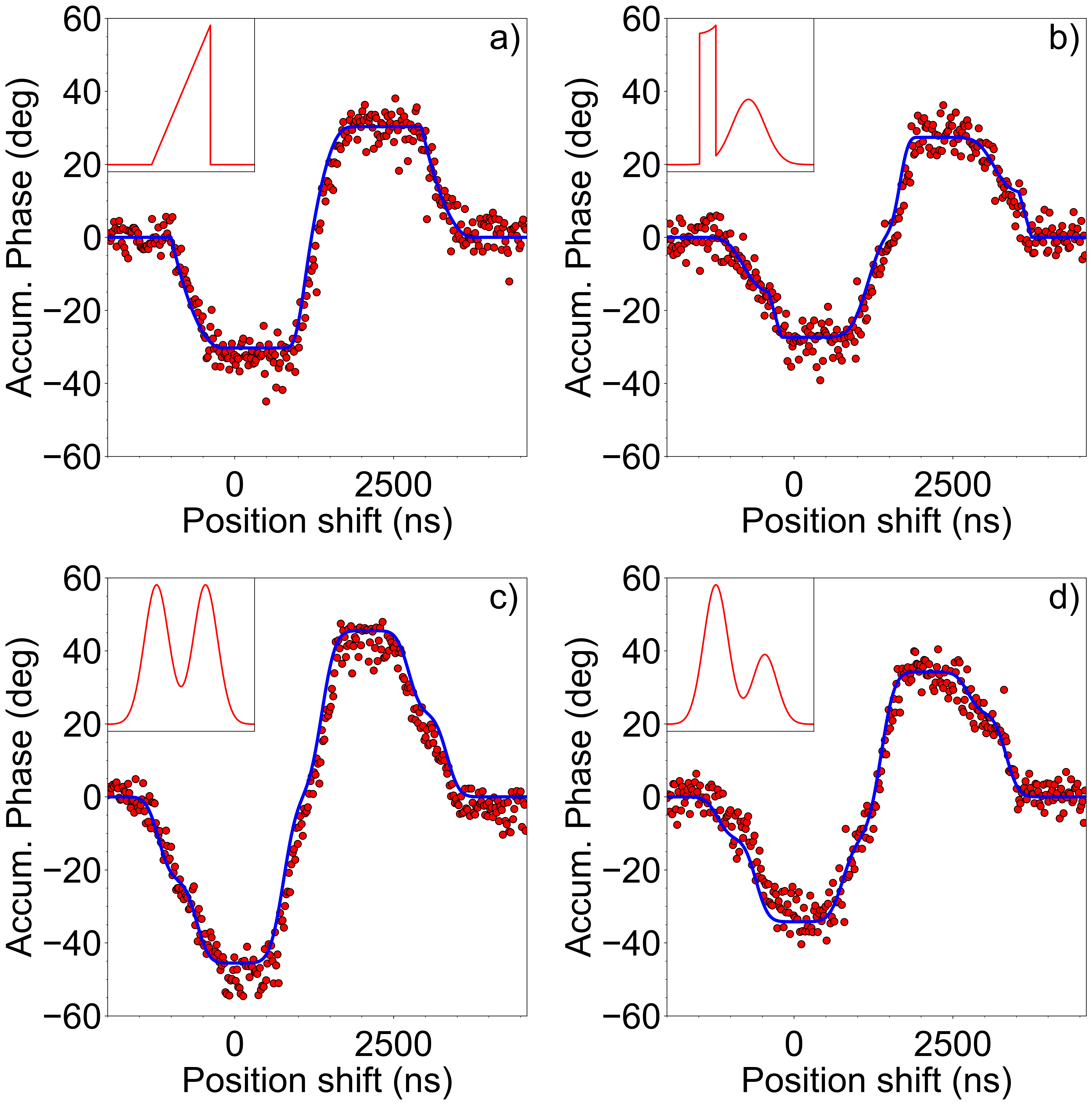}%
\caption{Phase accumulation of the echo obtained with sequence 2 on VOPt(SOCPh)\textsubscript{4} for differently shaped external signals (sketched in inset): (a) Sawtooth, (b) Square Wave + Gaussian, (c) Double Gaussian, (d) Asymmetric Double Gaussian. The experimental data are shown as red dots. The solid blue lines represent calculated curves obtained with the corresponding external signal parameters in Eq. \ref{eq:phase2}.\label{fig:FIG4}}
\end{figure}

To investigate the effect of memory time, $T$\textsubscript{m}, we use a single-crystal of magnetically diluted VOPt(SOCPh)\textsubscript{4}. Experiments are carried out at a fixed static field by applying Gaussian and rectangular-shaped signals. The resulting phase accumulation obtained with Sequence 2 (Fig. \ref{fig:FIG3}), shows that spins are sensitive to changes in all different Gaussian parameters ($B$\textsubscript{1,max},  $t$\textsubscript0, $\sigma$). It is also evident that narrower Gaussian signals lead to sharper edges in the phase accumulation curves (Fig. \ref{fig:FIG3} (c, g)). This is even more evident when the amplitude of the rectangular signal is changed (Fig. \ref{fig:FIG3} (d, h)).\\
The relatively long memory time (see Sec. \ref{sec:Exp}.C and Supplementary Information \cite{supp}) allows us to further test signals of increased complexity using Sequence 2 (Fig. \ref{fig:FIG4}). To this end, we use the copper coil to generate different field shapes, in particular: a single sawtooth, the sum of a rectangular and a Gaussian signal, and two consecutive Gaussian signals. The phase accumulation changes with the different excitations sent into the coil. As expected, the general behaviour for the phase accumulation is common for all signals since $\phi$\textsubscript{echo}($T$\textsubscript{seq},$s$) changes sign when the $\pi$ pulse is crossed. However, the shape of the curves and the maximum (and minimum) amount of the accumulated phase vary depending on the shape of the external signal. Additional differences can also be found in the slopes and features of the phase accumulation curves. Moreover, the behavior of the phase accumulation calculated with Eq. \ref{eq:phase2} by using the known signal parameters is in excellent agreement with experimental data (Fig. \ref{fig:FIG4}). 

%%%%%%%%%%%%%%%%%%%%%%%%
%%%%%%%%%%%%%%%%%%%%%%%%
%%%%%%%%%%%%%%%%%%%%%%%%%%%%%%%%%%%%%%%%%%%%%%%%%%%%%%%%%%%%%%%%%%%%%%%%%%%%%%%%%%%%%%%%%%%%%%%%%%%%%
\section{Discussion}

\begin{table*}
\centering
\begin{adjustbox}{width=\textwidth}
\begin{tabular}{c c c c c c c c c c c} 
\hline
\hline
   Sequence& \multicolumn{2}{ c }{$S$} & \multicolumn{2}{ c }{$S$ (l.b.)} & \multicolumn{2}{ c }{$S$\textsubscript{vol}}& \multicolumn{2}{ c }{$S$\textsubscript{vol} (l.b.)} & \multicolumn{2}{ c }{$A$\textsubscript{min}} \\ [0.5ex] 
   & \multicolumn{2}{ c }{($10^{-7}$ T Hz\textsuperscript{$-\frac{1}{2}$})}  & \multicolumn{2}{ c }{($10^{-8}$ T Hz\textsuperscript{$-\frac{1}{2}$})} & \multicolumn{2}{ c }{($10^{-11}$ T Hz\textsuperscript{$-\frac{1}{2}$} $\mu$m\textsuperscript{$\frac{3}{2}$})}  & \multicolumn{2}{ c }{($10^{-12}$ T Hz\textsuperscript{$-\frac{1}{2}$} $\mu$m\textsuperscript{$\frac{3}{2}$})} & \multicolumn{2}{ c }{($10^{-10}$ T s)}\\ 
 \cline{2-11}
  &VO(TPP)&VOPt(SOCPh)\textsubscript{4}& VO(TPP)&VOPt(SOCPh)\textsubscript{4}&VO(TPP)&VOPt(SOCPh)\textsubscript{4}&VO(TPP)&VOPt(SOCPh)\textsubscript{4}&VO(TPP)&VOPt(SOCPh)\textsubscript{4}\\ 
\hline
 \textcolor{black}{Seq. 1}  & 2.57 & \textcolor{black}{-} & 2.87 & \textcolor{black}{-} & 5.35 & \textcolor{black}{-} & 5.98 & \textcolor{black}{-} & 1.62 & \textcolor{black}{1.16}\\ 
 \textcolor{black}{Seq. 2} & \textcolor{black}{3.98} & 1.54 & \textcolor{black}{3.12} & 2.72 & \textcolor{black}{8.30} & 4.89 & \textcolor{black}{6.51} & 8.63 & \textcolor{black}{1.61} & 1.10\\
\hline
\hline
\end{tabular}
\end{adjustbox}
\caption{Summary of the sensitivity, $S$, of the concentration sensitivity, $S$\textsubscript{vol}, and of their lower bounds (l.b.) evaluated for Sequence (Seq.) 1 and 2. The values of the minimum area, $A$\textsubscript{min}, are also reported for both the VO(TPP) ($\rho$ = 2.3 $\cdot 10^{19}$ \cite{yamabayashiJACS2018, drewINORGCHIMACTA1984}) and the VOPt(SOCPh)\textsubscript{4} ($\rho$ = 1.0 $\cdot 10^{19}$ \cite{belliniThesis} ) samples}.
\label{tab:TABLE1}
\end{table*}

\subsection{Sensitivity}

The sensitivity of the two sequences can be estimated following the definition given in \cite{blankJMR2017}, that is: 
\begin{equation}
    S = \frac{B\textsubscript{min}}{\frac{S}{N} \sqrt{BW}},
    \label{eq:sensitivity}
\end{equation}
where $B$\textsubscript{min} is the minimum field that can be detected, $\frac{S}{N}$ is the signal-to-noise ratio and $BW$ is the bandwidth of the sequence. \\
$B$\textsubscript{min} can be estimated starting from the standard deviation of the phase accumulation $\sigma_{\phi}$ without an applied RF signal (see Supplementary Information \cite{supp}). This corresponds to the minimum phase change which can be detected under our experimental conditions. This phase value is then converted into a magnetic field using the corresponding transduction coefficient $\frac{d \phi}{d B}$, describing the response of the spin sensor (i.e., the phase variation for unit of applied magnetic field). The transduction coefficient is extracted from preliminary calibrations (see Supplementary Information \cite{supp}). 
Using the above definitions, we find $\sigma_{\phi} \approx 10.4$° and $\sigma_{\phi} \approx 8.3$°
for Sequence 1 and 2, respectively, for VO(TPP), and $\sigma_{\phi} \approx 4.4$° using Sequence 2 for VOPt(SOCPh)\textsubscript4  (see Supplementary Information \cite{supp}). For VO(TPP) we obtain $B$\textsubscript{min} = 1.14 $\cdot 10^{-4 }$ T and 1.64 $\cdot 10^{-4 }$ T for Sequence 1 and 2, respectively, while for VOPt(SOCPh)\textsubscript{4} the obtained value is $B$\textsubscript{min} = 4.78 $\cdot 10^{-5 }$ T for Sequence 2.
The signal-to-noise ratio is obtained by taking the amplitude of the signal at the first step of the sequence and dividing it by (see Supplementary Information \cite{supp}). The bandwidth is given by the inverse of the sequence repetition time, resulting in $BW$ = 50 Hz (see Supplementary Information \cite{supp}).  With the above values, the sensitivity obtained for VO(TPP) is $S$ = 2.57 $\cdot 10^{-7}$ T Hz\textsuperscript{$-\frac{1}{2}$} for sequence 1 and $S$ = 3.98 $\cdot 10^{-7}$ T Hz\textsuperscript{$-\frac{1}{2}$} for Sequence 2. For VOPt(SOCPh)\textsubscript{4}, the calculated sensitivity is $S$ = 1.54 $\cdot 10^{-7}$  for Sequence 2. The sensitivity values are summarized in Tab. \ref{tab:TABLE1} along with their lower bounds, which are \textcolor{black}{ based on} the Allan variance method \cite{Draganova2014,marinovAnalysisSensorsNoise2019}. This statistical analysis, used to extract the noise floor of sensors, is implemented by calculating the grouped standard deviation of a set of several identical repeated measurements. The lower bounds, derived using the values of $B$\textsubscript{min} estimated from the Allan variance, are reported in Tab. \ref{tab:TABLE1}.
Concentration sensitivity is then used to take into account the spin density ($\rho$) of the samples:

\begin{equation}
    S\textsubscript{vol} =  \frac{S}{\sqrt{\rho}}.
    \label{eq:sensitivity2}
\end{equation}
The resulting values \textcolor{black}{are} also collected in Tab. \ref{tab:TABLE1}.

\subsection{Minimum detectable area ($A$\textsubscript{min}) and Memory time ($T$\textsubscript{m})}

\begin{figure}
\includegraphics[width=0.95\columnwidth]{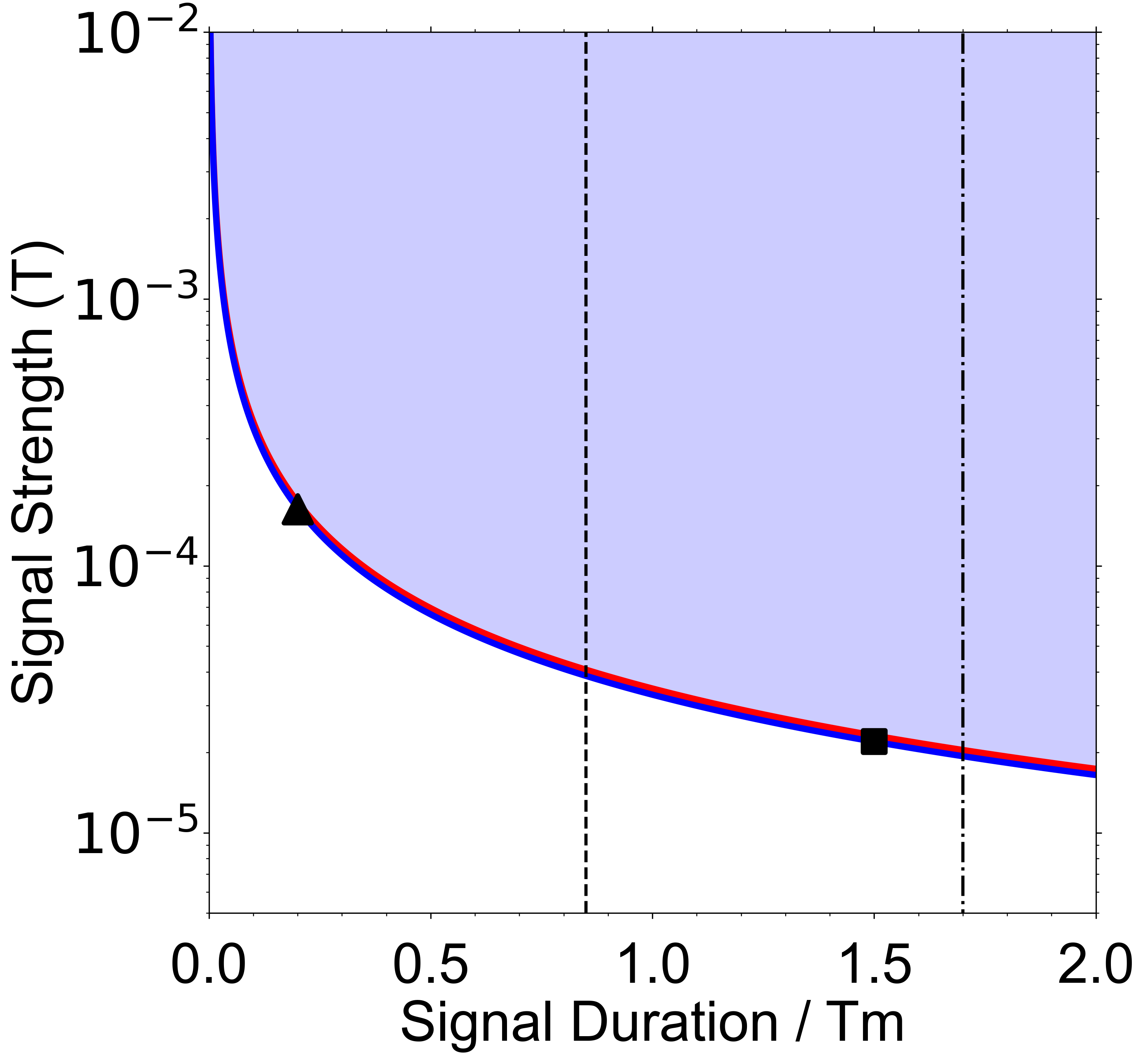}%
\caption{Detectable areas (blue shaded) that can be measured using Sequence 1 and Sequence 2 depending on the duration of the time-domain signal and on its field strength. The solid curves represent the  the minimum measurable area with the VOPt(SOCPh)\textsubscript{4} sample and using a Gaussian signal for Sequence 1 (red line) and for Sequence 2 (blue line). The signal duration is taken as $3\sigma$ (with $\sigma$ being the Gaussian width in Eq. \ref{eq:Gaussian}). The limit in signal duration ($T$\textsubscript{lim}) imposed by the memory time of the sample is shown by vertical lines for Sequence 1 (dashed line) and Sequence 2 (dashed-dotted line). The markers correspond to the Gaussian signal generated by the dipolar field of a $S$ = 1/2 spin located at a distance $d$ from the sensor and undergoing a full rotation (triangle: $\sigma$ = 130 ns, $d$ = 1.8 nm; square: $\sigma$ = 1.6 $\mu$s, $d$ = 4.4 nm).} 
\label{fig:FIG5}
\end{figure}

As described by Eq. \ref{eq:phase2}, the phase accumulation depends on the time integral of the signal (that is the area, $A$, under the field strength curve). This implies that the achievable sensitivity is \textcolor{black}{determined} by the interplay between a given signal strength and its corresponding duration, in other words, by a minimum detectable area, $A$\textsubscript{min}. This can be calculated for Sequences 1 and 2 from Eq. \ref{eq:phase2} using the phase noise of the echo when no external signal is applied (see Supplementary Information \cite{supp}). The obtained $A$\textsubscript{min} values are summarized in Tab. \ref{tab:TABLE1}. Results are further shown in Fig. \ref{fig:FIG5}, where the solid curves represent the $A$\textsubscript{min} limit for a Gaussian signal measured with VOPt(SOCPh)\textsubscript{4} using both sequences.\\
The upper limit for the time scale is imposed by the phase memory time of the sample. In fact, $T$\textsubscript{m} restricts the maximum duration of the external signal which can be measured, because the echo precession phase is lost for signals significantly longer than $T$\textsubscript{m} (and $\tau$). This is shown in Fig. \ref{fig:FIG5} as vertical dashed and dashed-dotted lines for Sequence 1 and Sequence 2 ($T$\textsubscript{lim}), respectively. These values are extracted from the echo phase measurement obtained during a Hahn echo sequence (see Supplementary Information \cite{supp}). 
Another limiting factor is the total area of the external signal, because the amplitude of the spin echo decreases as a function of the area and this can result in a total loss of the echo signal for too large values (see Supplementary Information \cite{supp}).
For Sequence 2 the limit for recording a full phase accumulation curve is twice as large as that in Sequence 1. This is due to the fact that $\tau$ remains fixed during the protocol. Here, we notice that for signals longer than $T$\textsubscript{lim} it is still possible to obtain a phase accumulation curve by using $\tau$ shorter than the signal duration (see Supplementary Information \cite{supp}). \\
To figure out some practical applications of our sensing protocols, we envisage a dimer containing a spin label (e.g. one of our quantum sensors) and a generic chemical analyte with a magnetic moment at a distance $d$. This dimer is diluted in a non-magnetic matrix.
Let's assume that we can induce a full spin rotation process on the analyte from the spin value $\hbar m_s$ to $-\hbar m_s$ and back to $\hbar m_s$, giving a transient signal at the sensor site. Although more specific profiles can be considered, here we assume a gaussian-shaped transient, only to refer to the signal in Eq. \ref{eq:Gaussian} and we neglect the decrease of the memory time of our spin sensor during its rotation. Looking at the plot in Fig. \ref{fig:FIG5}, a short signal of 0.1 $T$\textsubscript{m} duration could be detected if its field strength is at least 330 $\mu$T (triangle symbol). This corresponds to the dipolar field generated by a spin $S$ = 1/2 (e.g. a radical) at $d \approx 1.8$ nm from the molecular sensor or by a spin $S$ = 10 (for instance a Mn\textsubscript{12} or Fe\textsubscript8 Single Molecule Magnet) at $d \approx 3.8$ nm (always under the assumption $g$ = 2). For a long signal of 1.5 $T$\textsubscript{m} duration, the minimum strength of the Gaussian signal should be $\approx$ 22 $\mu$T (square symbol). This value corresponds to the dipolar field generated by a spin $S$ = 1/2 at $d \approx 4.4$ nm or by a spin $S$ = 10 at $d \approx 9.4$ nm.
Similarly, one may envision an application of our protocols in Metal Organic Framework (MOF) architectures recently proposed to host both quantum sensors and analytes \cite{sunRoomTemperatureQuantitativeQuantum2022}. In this case one should consider the mean field generated by a collection of analytes close to the sensor’s site and collectively triggered at the same time. These values highlight that Fig. \ref{fig:FIG5} may help to evaluate the feasibility of quantum sensing experiments in different realistic situations. 

%%%%%%%%%%%%%%%%%%%%%%%%
%%%%%%%%%%%%%%%%%%%%%%%%

%%%%%%%%%%%%%%%%%%%%%%%%%%%%%%%%%%%%%%%%%%%%%%%%%%%%%%%%%%%%%%%%%%%%%%%%%%%%%%%%%%%%%%%%%%%%%%%%%%%%%
\section{Conclusions and outlooks}

We developed two quantum sensing protocols for the detection of \textcolor{black}{time-dependent} magnetic fields by using molecular spins as detectors. These protocols \textcolor{black}{expand} the sensing capabilities of molecular spins beyond quantum sensing of AC magnetic fields \cite{bonizzoniQuantumSensingMagnetic2024}. \textcolor{black}{Both our sequences} have sensitivities of the order of few $S \approx 10^{-7}$ T Hz\textsuperscript{$-\frac{1}{2}$} with lower bounds approaching few $S \approx 10^{-8}$ T Hz\textsuperscript{$-\frac{1}{2}$}. 
These values are in the same range as those \textcolor{black}{that} we previously obtained in AC quantum sensing experiments \cite{bonizzoniQuantumSensingMagnetic2024}. \textcolor{black}{Although experimental conditions, setups, protocols used and techniques for sensor readout prevent from a direct comparison, we just mention that sensitivity values in the $10^{-6} - 10^{-9}$ T Hz\textsuperscript{$-\frac{1}{2}$} range have been reported for time-dependent detection and reconstruction of arbitrary signals in \cite{zopesReconstructionfreeQuantumSensing2019a,balasubramanianNATMAT2009} for NV centers.}\\
A common feature of \textcolor{black}{our} sequences is the minimum area of the signal which can be measured. The values are $A$\textsubscript{min} $\approx$ 1.6 $\cdot 10^{-10}$ T s for VO(TPP) and $A$\textsubscript{min} $\approx$ 1.1 $\cdot 10^{-10}$  T s for VOPt(SOCPh)\textsubscript{4}. These values allow for a good trade-off between the required signal duration and the available sensitivity. \textcolor{black}{In addition, Sequence 2 can be used even when the delay of the external signal with respect to the trigger is unknown. In fact, since the interpulse delay is fixed, the position shift, $s$, can be changed until $B$\textsubscript1($t$,$s$) starts to cross the MW sequence. For these reasons, Sequence 2 results more versatile and more robust in the context of time-dependent signals detection.}\\ 
\textcolor{black}{As possible future development of our protocols, we note that Sequence 2 could in principle be used to perform a full reconstruction of a time-dependent signal (see Supplementary Information \cite{supp}), similarly to what reported in \cite{zopesReconstructionfreeQuantumSensing2019a}. This, however, requires an improvement of the signal-to-noise ratio and of the homogeneity of the MW field, achievable with an optimization of the experimental setup.
Another possible outlook relies on using machine learning protocols, which proved to be a successful tool for phase prediction in one
of our previous works \cite{bonizzoniMachineLearningAssistedManipulationReadout2022}, to assist the detection of the signal and, eventually, its reconstruction.}

\section*{Acknowledgements}
The authors acknowledge the contributions of Mr. Giacomo Bellini (UniMORE), who participated in the preparation and investigation of samples, and Prof. Roberta Sessoli (UniFi) for critical revision of the manuscript.
This work was partially supported by the project SMILE-SQUIP funded by the Italian NQSTI - National Quantum Science and Technology Institute code PE0000 0023 and by the U.S. Office of Naval Research Award No. N62909-23-1-2079.
European Research Council partially funded this work through the ERC SYNERGY project CASTLE (no. 101071533).

\bibliography{bibliography}



\section*{Supplementary material}

# SUPPLEMENTARY: Quantum Sensing of time-dependent magnetic signals with molecular spins

Matteo Lanza,[1,2,*] Claudio Bonizzoni,[1,2] Olga Mironova,[3] Fabio Santanni,[4] Alessio Nicolini,[3] Alberto Ghirri,[2] Andrea Cornia,[3] and Marco Affronte[1,2]

[1] *Dipartimento di Scienze Fisiche Informatiche e Matematiche,*
*Università degli Studi di Modena e Reggio Emilia, via G. Campi 213/A, 41125 Modena, Italy*
[2] *Istituto Nanoscienze CNR, Centro S3, via G. Campi 213/A, 41125 Modena, Italy*
[3] *Dipartimento di Scienze Chimiche e Geologiche e UdR INSTM,*
*Università degli Studi di Modena e Reggio Emilia, via G. Campi 103, 41125 Modena, Italy*
[4] *Dipartimento di Chimica "U. Schiff", Università di Firenze,*
*Via della Lastruccia 3, Sesto F.no, 50019, Italy*
(Dated: November 11, 2025)

# INTRODUCTION

This supplementary information contains details on the model used for the description of the phase accumulation (Section I) and on the parameters used for the microwave Sequences (Section II). The preparation and orientation of the samples, and the calibration of the spin sensors can be found throughout Sections III, IV, V and VI. A detailed description of the Allan Variance method used for the estimation of the minimum detectable field appears in Section VII. The effect of the memory time $T_\mathrm{m}$ on both Sequences is discussed in Section VIII and IX. Finally, the results of fitting the experimental data with the phase accumulation model discussed in Section I can be found in Section X.

# CONTENTS

---

* matteo.lanza@unimore.it

## I. PHASE ACCUMULATION MODEL

The phase accumulation for both Sequence 1 and 2 depends on the area under the external signal, $B_1(t,s)$, during the sequence duration [1]. Under the assumption that no phase is accumulated during the MW manipulating pulses, the phase accumulation equation can be written as:

$$\phi_{\text{echo}}(T_{\text{seq}}, s) = \int_{t_{\frac{\pi}{2},\text{end}}}^{t_{\pi,\text{start}}} \gamma B_1(t,s)dt - \int_{t_{\pi,\text{end}}}^{t_{\text{echo}}} \gamma B_1(t,s)dt, \quad (S1)$$

This approximation holds in a regime in which the length of the external signal is longer with respect to the pulse duration or it varies slowly in time. Conversely, when $B_1(t, s)$ shows fast variations in time or it is shorter than the $\pi$ pulse, this approximation fails to describe correctly the phase accumulation behaviour (see Fig. S1). In this case the rotation of the spins on the Bloch sphere is taken into account by including a third term in the integral as [2]:

$$\phi_{\text{echo}}(T_{\text{seq}}, s) = \int_{t_{\frac{\pi}{2},\text{end}}}^{t_{\pi,\text{start}}} \gamma B_1(t,s)dt - \int_{t_{\pi,\text{start}}}^{t_{\pi,\text{end}}} \gamma B_1(t,s) \sin\left(\frac{t - t_{\pi,\text{start}} - \frac{T_\pi}{2}}{T_\pi/\pi}\right) dt - \int_{t_{\pi,\text{end}}}^{t_{\text{echo}}} \gamma B_1(t,s)dt. \quad (S2)$$

As described in the main text, in Eq. S1 and S2 $t_{\frac{\pi}{2},\text{end}}$ is the time at which the $\pi/2$ ends, while $t_{\pi,\text{start}}$ and $t_{\pi,\text{end}}$ are the time at which the $\pi$ pulse begins and ends. $t_{\text{echo}}$ is the time at which the echo is observed and $s$ is the position of the RF signal. All times are referred to the beginning of the MW sequence and, hereafter we will fix $t_{\frac{\pi}{2},\text{start}} = 0$. We check the validity of Eq. S1 and S2 under our experimental conditions by measuring the phase accumulation for different values of the ratio, $r$, between the duration of a square wave external signal and the duration of the $\pi$ pulse. The results are shown in Fig. S1. It is possible to see that for high values of $r$ the two calculated curves (blue and green, corresponding to Eq. S2 and Eq. S1, respectively) are in good agreement, while for low values of $r$ Eq. S1 fails to reproduce the experimental data. Due to the results above we choose to use Eq. S2 to analyze our experimental data and not the approximated version given by S1.

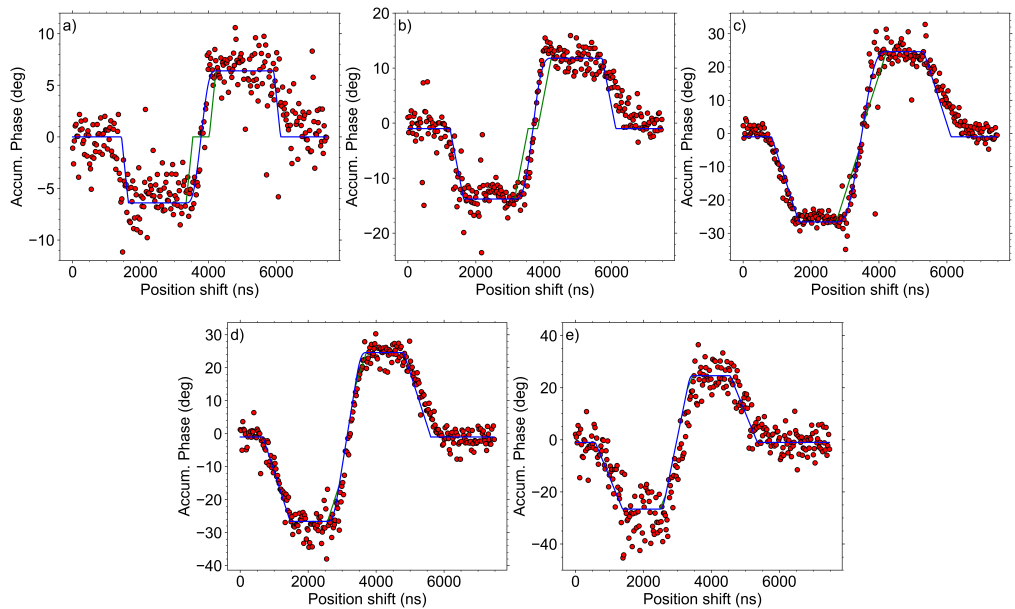

FIG. S1. Phase accumulation curves taken with Sequence 2 for increasing values of the ratio, $r$, between the duration of the external signal and the duration of the $\pi$ pulse. The experimental data are shown as red dots, while the solid lines represent the calculated curves obtained inserting the known external signal parameters in S2 (blue) and in S1 (green). The values used are: $r = 0.29$ (a), $r = 0.59$ (b), $r = 1.18$ (c), $r = 2.35$ (d) and $r = 4.71$ (e).

## II. MW PULSE SEQUENCES

The data reported in this work for VO(TPP) are obtained with MW pulse durations $T_{\frac{\pi}{2}} = 90$ ns, $T_\pi = 180$ ns for Sequence 1 and $T_{\frac{\pi}{2}} = 150$ ns and with $T_\pi = 300$ ns, for Sequence 2, respectively. The interpulse delays used are $\tau =$

1100 ns for Sequence 1 and $\tau$ = 1200 ns for Sequence 2.
The MW pulse durations for the VOPt(SOCPh)$_4$ tests are $t_{\frac{\pi}{2}}$ = 170 ns and with $t_\pi$ = 340 ns for both Sequence 1 and 2. The interpulse delays used for this sample is $\tau$ = 1800 ns for both sequences. A relaxation time $t_{\text{relax}}$ = 20 ms is added at the end of each sequence repetition to avoid sample saturation. The duration of the pulses for the Hahn echo sequence are set after preliminary calibrations. Each acquisition results from the average of 1000 individual traces (exception made for Sequence 1 test on VO(TPP) in which the number of averages is increased to 2000).

### III. SAMPLES PREPARATION AND ORIENTATION

The first sample is a crystalline solid solution of VO(TPP) (2 mol%) in TiO(TPP) (TPP = tetraphenylporphyrin), prepared according to a literature procedure [3]. Due to the powder form of this sample, no orientation on the YBCO resonator is needed. The second sample was obtained by incorporating the lantern-like complex [VOPt(SOCPh)$_4$] [4, 5] (1 mol%) in crystalline [TiOPt(SOCPh)$_4$] · 2THF [6]. This diamagnetic host was prepared by reacting TiCl$_2$(acac)$_2$ in MeOH with a water solution of [Pt(SOCPh)$_4$]$^{2-}$, followed by isolation of the product and recrystallization from THF to give yellow block-like tetragonal crystals ($I\bar{4}c2$ space group). The unit cell contains eight molecules lying with their Ti-Pt vectors perpendicular to the fourfold axis and arranged in four staggered dimers [TiOPt(SOCPh)$_4$]$_2$. Magnetically diluted samples were prepared by co-crystallization of the mixed vanadyl and titanyl complexes. Powder X-ray diffraction proved that at doping levels from 10 to 0.1 mol% they are isomorphous to the pure titanyl matrix. A single crystal containing 1 mol% of [VOPt(SOCPh)$_4$] was manually oriented to set the magnetic field parallel to the tetragonal axis and, consequently, perpendicular to the Ti-Pt direction of all molecules.

### IV. CALIBRATION OF THE MOLECULAR SPIN SENSORS

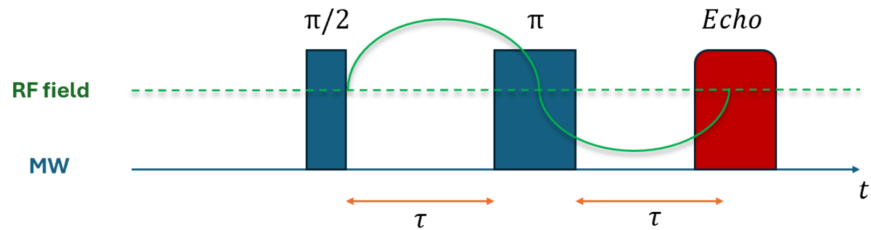

FIG. S2. Pulse sequence used to calibrate the phase response of the spin ensemble to the external magnetic field. A Hahn echo sequence with a fixed $\tau$ is used to produce a spin echo while a sinusoidal RF field is applied. The period of the sinusoidal signal is chosen to match the MW pulse duration.

We calibrate the response of our molecular spin sensors by using a sinusoidal AC field generated with the Cu coil and synchronized with the interpulse delay $\tau$. The AC field can be described as: $B_1(t,s=0) = B_1 sin(\frac{2\pi}{T}t)$, where $T$ = 2 $\tau$ is the period of the sinusoidal signal (see Fig. S2). The amplitude, $B_1$, of the AC field is increased step-by-step by changing the RF coil voltage while $T$ is kept fixed. For each $B_1$ value, the phase accumulated by the spin echo is recorded, similarly to what was previously done in [7]. Examples of calibration curves obtained for VO(TPP) and VOPt(SOCPh)$_4$ samples are shown in Fig. S3. A linear fit allows for the extraction of the transduction coefficient, $\frac{d\phi}{dB}$, of the molecular spin sensors. Under our experimental conditions we obtain $|\frac{d\phi}{dB_1}|$ = 9.07 · 10$^4$ deg/T for the VO(TPP) and $|\frac{d\phi}{dB_1}|$ = 9.19 · 10$^4$ deg/T for VOPt(SOCPh)$_4$. This calibration procedure is repeated every time a new measurement session is performed, since small changes in the sample or coil position can lead to variation in the corresponding $|\frac{d\phi}{dB_1}|$.

The RF coil used in this experiment is the same used in our previous work in which we performed a calibration procedure to extract the value of the field generated at its center [7]. The calibration curve we previously extracted is still valid, since the experiment is performed in the same conditions, and gives us $\frac{dB_1}{dV}$ = 756 $\mu$T/V.


4

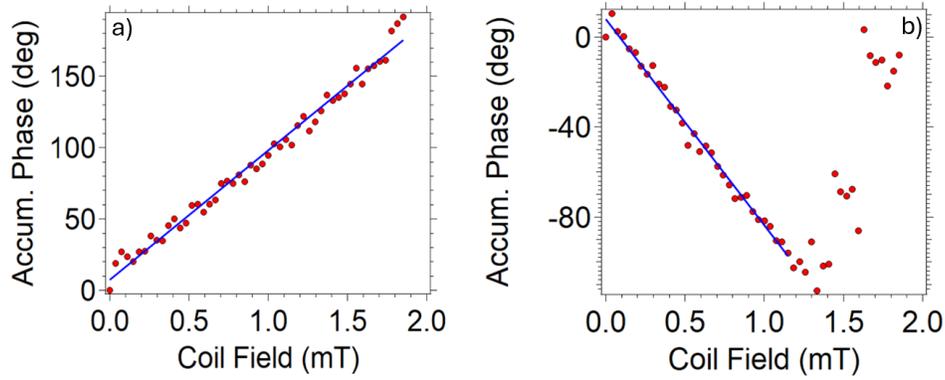

FIG. S3. Phase accumulation measured for the VO(TPP) sample (a) and VOPt(SOCPh)$_4$ (b). Blue lines correspond to linear fits, whose slope gives the transduction coefficient $\frac{d\phi}{dB_1}$.

## V. ESTIMATION OF THE TRANSDUCTION PARAMETER $\gamma$

During the estimation of $\gamma$, the position shift value is fixed to $s = 0$ as in Sec. IV. Therefore, in this section, the dependence of the coil field on $s$ will be omitted to simplify the notation to $B_1(t)$. The phase accumulation of a spin echo in an Hahn echo sequence under the effect of an external magnetic field signal, $B_1(t)$, is given by $\phi_{\text{echo}} = \gamma \int_0^{T_{\text{seq}}} B_1(t)\,dt$ [1], where $\gamma$ corresponds to the phase accumulated for unit area by the echo (or the phase accumulated for unit of field and time). Using the protocol in Fig. S2 and the phase inversion given by the $\pi$ pulse, it is possible to add the area from both half periods of $B_1(t)$ constructively. The total area of the RF signal over the whole MW sequence can then be written as:

$$\phi_{\text{echo}} = \int_0^{T_{\text{seq}}} \gamma B_1 \sin(\frac{2\pi}{T}t)dt = 2 \int_0^{\tau} \gamma B_1 \sin(\frac{2\pi}{T}t)dt = 2B_1\gamma \int_0^{\tau} \sin(\frac{2\pi}{T}t)dt = \frac{4B_1\tau\gamma}{\pi}. \tag{S3}$$

The $\gamma$ parameter can be retrieved by measuring the phase accumulation at a given field value and inverting the phase accumulation equation to obtain $\gamma = \frac{\pi \phi_{\text{echo}}}{4\tau B_1} = \frac{\pi}{4\tau}|\frac{d\phi}{dB_1}|$. Since $\gamma$ depends on $|\frac{d\phi}{dB_1}|$, also its value might change with the transduction coefficient after each new sample and coil positioning.
Using the data reported in Fig. S3 and in Sec. IV we obtain $\gamma = 6.476 \cdot 10^{10}$ deg T$^{-1}$ s$^{-1}$ for VO(TPP) and $\gamma = 4.010 \cdot 10^{10}$ deg T$^{-1}$ s$^{-1}$ for VOPt(SOCPh)$_4$. These values are used in conjunction with the known $B_1(t,s)$ parameters and Eq. 2 of the main paper to simulate the expected phase accumulation.

## VI. ESTIMATION OF MAGNETIC FIELD SENSITIVITY

The sensitivity of the sample defined in the main text depends on the following quantities: the minimum detectable field, $B_{\text{min}}$, the signal-to-noise ratio, $S/N$, and the bandwidth of the sequence, $BW$. The latter is defined as the inverse of the sequence duration, and due to the fixed relaxation time $t_{\text{relax}} = 20$ ms added at the end of each sequence, it is fixed at $BW = 1/20$ ms $= 50$ Hz for all sequences.
$B_{\text{min}}$ is extracted for each sequence and for each sample as the standard deviation of the phase signal when no external RF signal is applied to the coil (an example of this data is shown in Fig. S4). Any signal larger than this background noise can produce a detectable phase variation. For VO(TPP) and under our experimental conditions, this results in 10.48° for Sequence 1 and 8.34° for Sequence 2. For VOPt(SOCPh)$_4$, this results are 4.67° for sequence 1 and 4.47° for sequence 2. The corresponding values of $B_{\text{min}}$ can be found by dividing the standard deviation values by the corresponding $\frac{d\phi}{dB_1}$ parameter. The $S/N$ value can be obtained as the ratio of the amplitude of the echo signal over the standard deviation of the signal taken at a point where the echo is not present (as shown in Fig. S5). This results in $S/N = 63$ for Sequence 1 and $S/N = 58$ for Sequence 2 on VO(TPP). For the VOPt(SOCPh)$_4$ sample, we have $S/N = 44$ for Sequence 2. All the parameters used for estimating the sensitivity are collected in Tab. S1

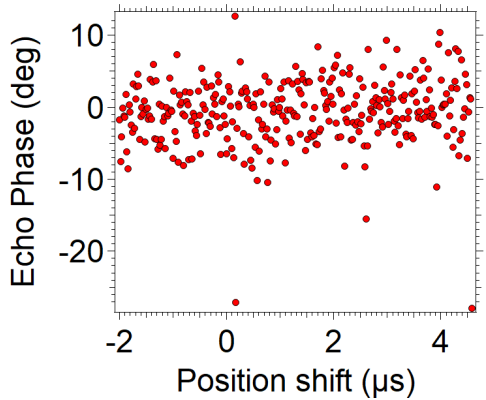

FIG. S4. Phase of the spin echo measured with Sequence 2 on the VOPt(SOCPh)$_4$ sample for $B_1(t) = 0$ mT. The standard deviation of the signal is taken as the minimum phase variation that should be produced by an external magnetic field in order to be measurable. This corresponds to 4.47° for the data in this panel.

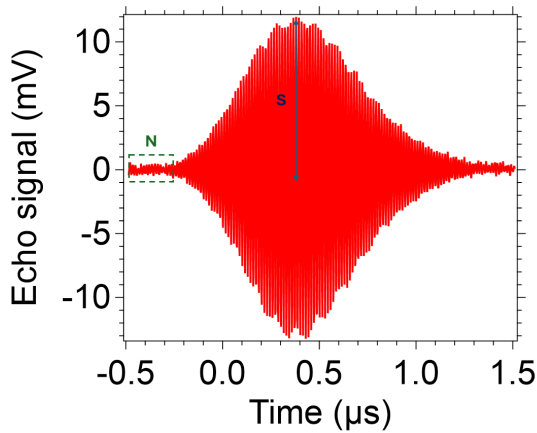

FIG. S5. Raw echo signal measured on oscilloscope on channel I of the output mixer for the VOPt(SOCPh)$_4$ sample. The signal-to-noise ratio can be extracted by dividing the amplitude of the echo (S) by the standard deviation of the signal in the area with no echo (N).

## VII. ESTIMATION OF MAGNETIC FIELD SENSITIVITY WITH THE ALLAN VARIANCE METHOD

An alternative and independent estimation of the minimum detectable field can be obtained with the Allan variance method [8, 9]. This statistical method allows us to determine the stability of the quantum sensor and the dominant type of noise. To estimate the Allan Variance, a set of $N$ repeated measurements under identical experimental conditions is taken and successively divided into $K$ clusters of $m = K/N$ points. For each cluster, the average value,

| Sequences | $BW$ (Hz) | $B_{min}$ ($10^{-5}$ T) | | $S/N$ | | $S$ ($10^{-7}$ T Hz$^{-\frac{1}{2}}$) | |
|---|---|---|---|---|---|---|---|
| | | VO(TPP) | VOPt(SOCPh)$_4$ | VO(TPP) | VOPt(SOCPh)$_4$ | VO(TPP) | VOPt(SOCPh)$_4$ |
| Sequence 1 | 50 | 11.45 | - | 63 | - | 2.57 | - |
| Sequence 2 | 50 | 16.43 | 4.78 | 58 | 44 | 3.98 | 1.54 |

TABLE S1. Summary of the magnetic field sensitivity, $S$, estimated for Sequence 1 and 2 on VO(TPP) and for Sequence 2 on VOPt(SOCPh)$_4$. All parameters required for the sensitivity estimation are also included in the table.

$\bar{a}_k$, is used to calculate the Allan variance, $\sigma_A{}^2(m)$, as:

$$\sigma_A{}^2(m) = \frac{1}{2(K-1)} \sum_{k=1}^{K-1} (\bar{a}_{k+1} - \bar{a}_k)^2 \qquad (S4)$$

The Allan deviation, $\sigma_A(m)$, can be obtained by taking the square root of $\sigma_A^2(m)$ and plotted as a function of $m$ in a log-log scale, as in Fig. S6. The error bars can be obtained from [8]:

$$\delta\sigma_A{}^2(m) = \frac{1}{2\sqrt{K-1}} \qquad (S5)$$

We evaluate the Allan variance under our experimental conditions using a Hahn echo sequence and a rectangular-shaped signal. We consider three different cases: i) external signal overlapping with the $\pi/2$ pulse, ii) external signal between the $\pi/2$ and $\pi$ pulse, and iii) external signal overlapping with the $\pi$ pulse. For each case, we use $N = 300$ consecutive and independent acquisitions and we obtain a corresponding set of echo phase values, from which Allan deviation is evaluated. The conversion of the Allan deviation from degrees into the corresponding field is obtained using the transduction coefficient, $\frac{d\phi}{dB_1}$. A linear fit of the Allan deviation curves gives a slope of $\sim$ -0.5 for all different regimes and for both samples. This value indicates that the dominant type of noise is white noise [9]. Taking the average of the last point from the three curves, a lower bound value for $B_{\min}$ can be obtained. These values are lower than the ones obtained in Sec. VI.

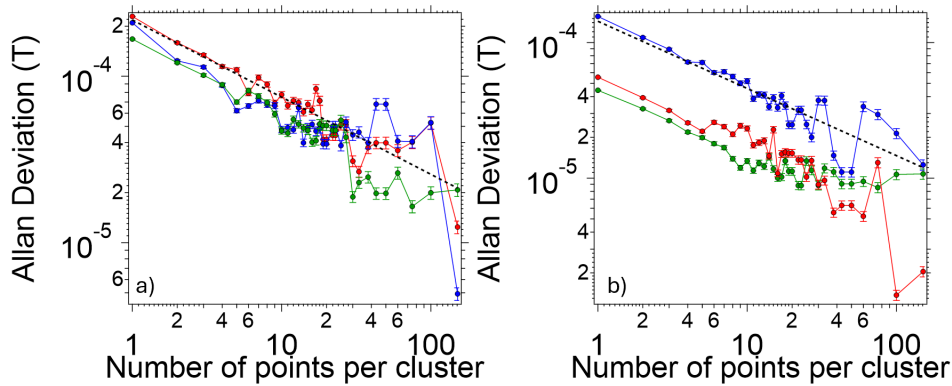

FIG. S6. Allan deviation plots taken for the VO(TPP) (a) and VOPt(SOCPh)$_4$ (b) samples. The curves data are taken by repeating an Hahn echo sequence $N = 300$ times in three different regimes: external signal overlapping with the $\pi/2$ pulse (red), external signal between the $\pi/2$ (blue) and $\pi$ pulse and external signal overlapping with the $\pi$ pulse (green). A linear fit (dashed black line) gives a slope of $\sim$ 0.5, corresponding to white noise.

## VIII. ROLE OF THE MEMORY TIME IN SEQUENCE 1

For Sequence 1, the memory time $T_m$ limits the duration of the external signal to be detected through the maximum $\tau$ which can be used. In fact, the phase accumulation is not visible for signals significantly longer than $T_m$ since the spin coherence is lost and a measurable echo is no longer visible. For Sequence 1, the use of a sample with longer memory times allows extending the length of the interpulse delay and, hence, the maximum signal duration. This value, denoted as $T_{lim}$, which is longer than $T_m$, can still give a measurable echo.

We characterize this value by performing a measure of the Hahn echo memory time of our samples, i.e. by recording the echo amplitude and phase as the interpulse delay $\tau$ is increased, as in Fig. S7 and Fig. S8. The echo amplitude shows an exponential decay for increasing interpulse delay, while the echo phase remains constant up to a value beyond which the phase (i.e., coherence) is lost. To phase memory time, $T_m$, is estimated by fitting the data with:

$$A_{Echo}(\tau) = A_{Echo}(0) e^{-\left(\frac{2\tau}{T_m}\right)^x}, \qquad (S6)$$

where $x$ is a stretching parameter. Since the data are normalized on the starting point $A_{Echo}(0)$ is fixed to 1.

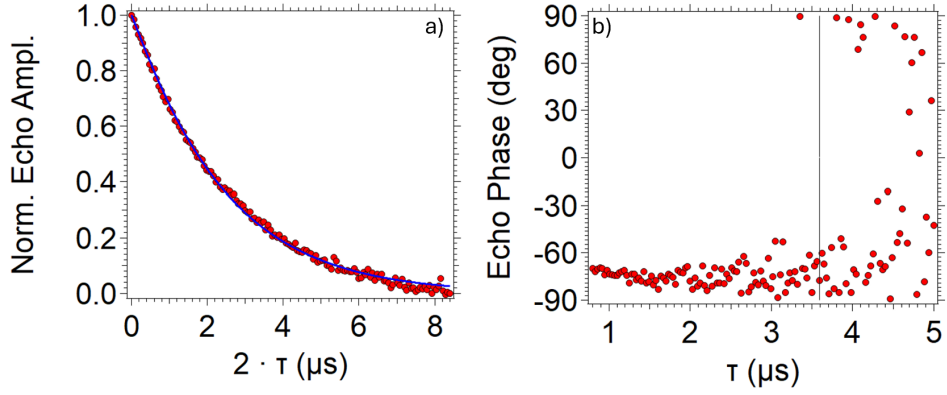

FIG. S7. Memory time for the VO(TPP) sample measured at 3 K and on the main powder line. a) Echo amplitude normalized on the intial value as a function of twice the interpulse delay. The starting value of $\tau$ is shifted to zero. The values for the memory time is $T_\mathrm{m} = 2.0 \pm 0.5$ $\mu$s and the stretching parameter $x = 1.0 \pm 0.1$ are extracted by fitting the data with Eq. S6 (blu line). b) Echo phase extracted during the same measurement in (a). The phase signal starts to heavily scatter around $2\tau \approx 3.6$ $\mu$s (vertical gray line) due to the echo being too small to be measured.

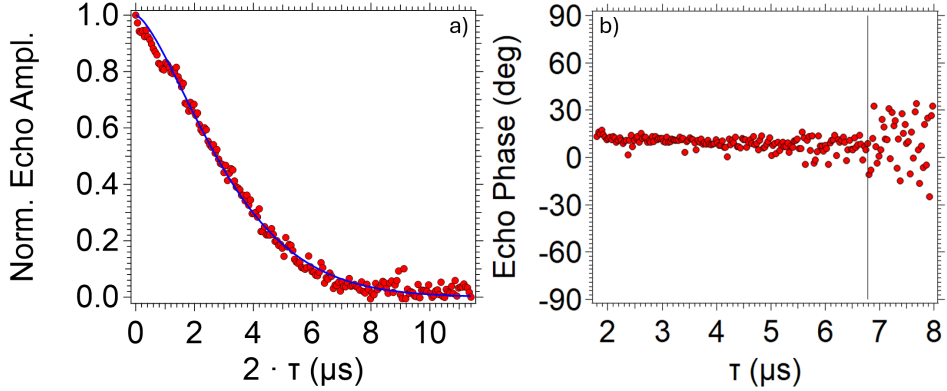

FIG. S8. Memory time for the VOPt(SOCPh)$_4$ sample measure at 3 K on the line giving the strongest signal. a) Echo amplitude normalized on the first value as a function of twice the interpulse delay. The starting value of $\tau$ is shifted to zero. Fit performed with Eq. S6 give $T_\mathrm{m} = 4.0 \pm 0.5$ $\mu$s and $x = 1.5 \pm 0.1$. b) Echo phase extracted during the same measurement shown in (a). The phase signal starts to heavily scatter around $2\tau \approx 6.8$ $\mu s$ (vertical gray line) due to the echo being too small to be measured.

Looking at the amplitude of the echo during the calibration procedure described in sec. IV (see Fig. S9 (a)), it is possible to see that the echo amplitude descreases as the external field strength is increased. The same behaviour can be observed for the amplitude of the echo during a Sequence 2 run (Fig. S9 (b)) in which the external signal strength is kept fixed and only the area inside the sequence is changing by incfreasing the position shift. This indicates that $T_\mathrm{lim}$ acts as an upper bound for the signal duration, since the echo could vanish before reaching this value, leading to a non measurable phase signal.

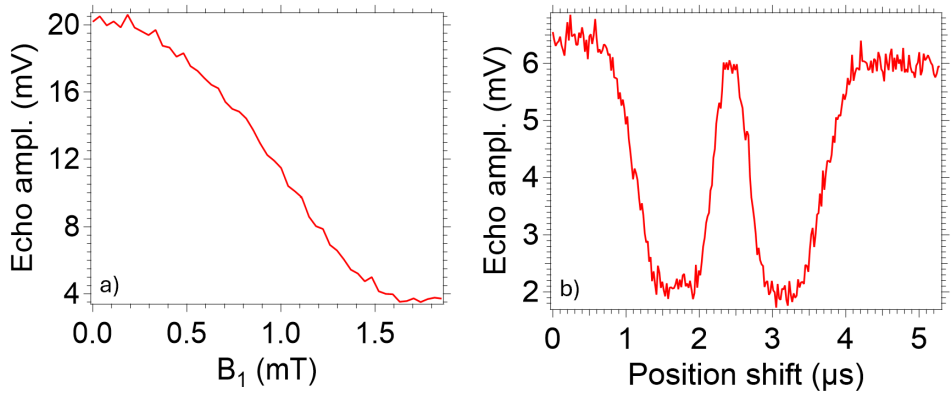

FIG. S9. (a) Echo amplitude for the VO(TPP) sample as the external field strength is increased during the sample calibration procedure. (b) Echo amplitude as the area of the external signal is changed at fixed strength with Sequence 2 (b).

To test these effects and the role of $T_m$, both the $VOPt(SOCPh)_4$ and VO(TPP) have been used with Sequence 1 to measure rectangular wave signals of long duration (1 - 3 $\mu s$) with respect to $\tau$. The results show that with the VO(TPP) sample cannot completely measure the phase accumulation of a signal of $1\mu s$ duration. With the VOPt(SOCPh)$_4$ instead, it is possible to see the full phase accumulation for a 2 $\mu s$ long signal by sweeping $\tau$ up to 5500 ns (see Fig. S10). For the 3 $\mu s$ signal duration, the echo signal is lost under the noise baseline for $\tau \approx 6.5$ $\mu s$ (see Fig. S10).

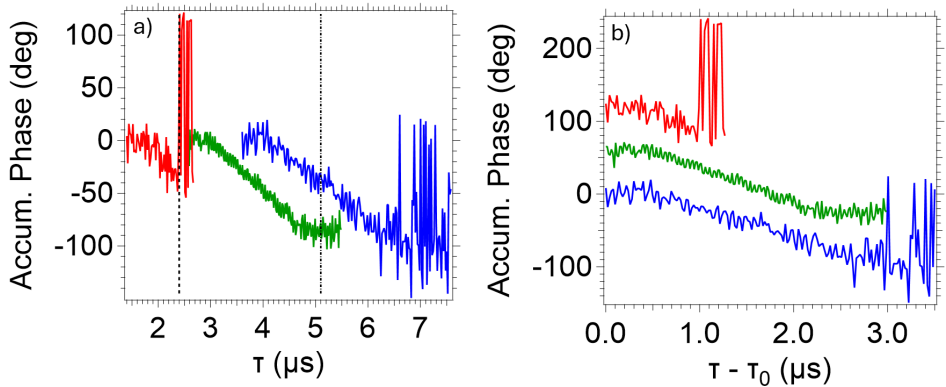

FIG. S10. a) Phase accumulation obtained with Sequence 1 when measuring an external rectangular field signal. For VO(TPP), the signal duration is 1 $\mu s$ (red curve), while for VOPt(SOCPh)$_4$ a duration of 2 $\mu s$ (green) or of 3 $\mu s$ (blue) is used. For all traces, the phase coherence get lost after a certain $\tau$ value which depends on both $T_\mathrm{m}$ and on the total area of phase accumulation. Vertical lines show the point where $\tau = T_\mathrm{m}$ for the VO(TPP) (dashed) and VOPt(SOCPh)$_4$ (dashed-dotted) samples; with values obtained from the fits of Fig. S7, S8. In (b) the phase accumualation curves are shifted to zero for better comparison (a vertical offset is added for clarity).

## IX. ROLE OF THE MEMORY TIME IN SEQUENCE 2

$T_\mathrm{m}$ does not act as a limit for the duration of the external signal on Sequence 2 since the $\tau$ value is kept constant during the measurement. This holds as long as $\tau$ is not too long to prevent the observation of spin coherence. With Sequence 2, the phase accumulation can still be visible even for signals that are longer than $\tau$. However, in this latter case, the phase accumulation does not show any plateau region/s, as in Fig. S11, since the area of the signal can no longer be completely included between the two pulses.

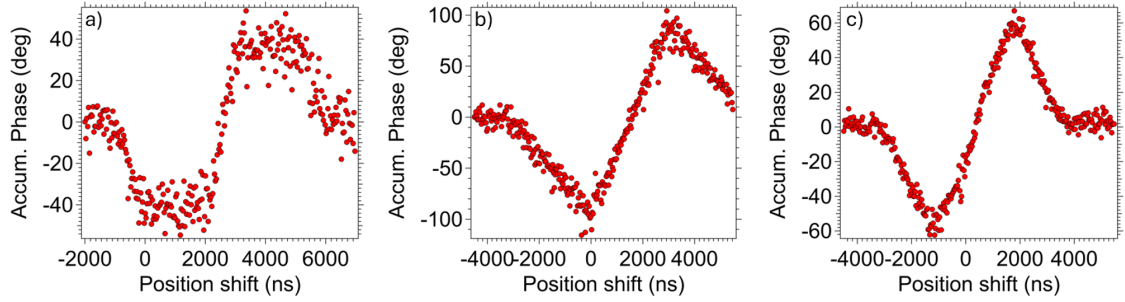

FIG. S11. Phase accumulation curves taken with Sequence 2 for increasing values of the ratio, $r$, between the duration of the external signal and the duration of $\tau$. The values used : $r = 0.26$ (a), $r = 1$ (b) and $r = 1.67$ (c). The plateau disappears for $r \geq 1$.

## X. FIT OF $B_1(t, s)$ PARAMETERS FROM EXPERIMENTAL DATA

The phase accumulation in Eq. S2 can be used to fit the experimental data shown in the main paper. We test this possibility by fitting the data obtained for VOPt(SOCPh)$_4$ using Sequence 2 and a Gaussian signal. To this end, we use a Gaussian signal in the form $B_1(t, s) = B_{1,\max} \, e^{\left(-\frac{(t-t_0-s)^2}{2\sigma^2}\right)}$ into S2 and we treat $B_{1,\max}$, $t_0$, $\sigma$ as fit paramters. An example of the signal constructed with the obtained fit parameters is shown in Fig. S12.
Tab. S2 at the end of the section contains the summarized fit parameters obtained for different tested Gaussian signals.

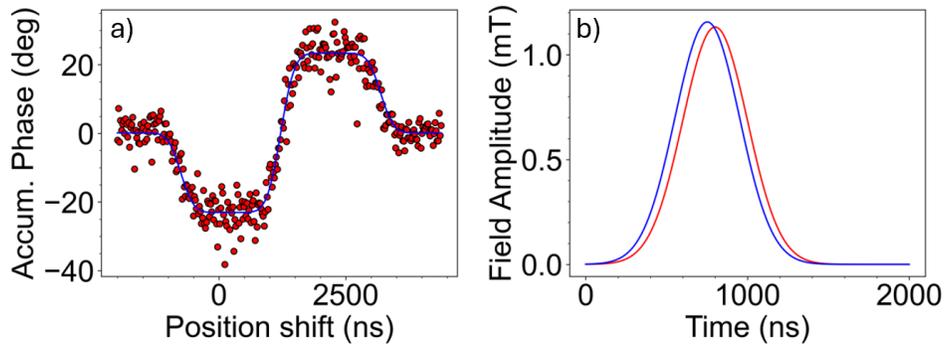

FIG. S12. a) An example of the fit of the phase accumulation for a Gaussian signal $B_1(t, s)$ (real parameters: $B_{1,\max} = 1.13$ mT, $t_0 = 800$ ns, $\sigma = 200$ ns) obtained with Sequence 2 and the VOPt(SOCPh)$_4$ sample. The experimental data (red dots) are fitted with Eq. S2 using a Gaussian model (blue line). b) The real shape of $B_1(t, s)$ (red) compared with the one calculated with Eq. S2 using the fit parameter obtained in (a).

| Data | $B_0$ (mT) | | $x_0$ (ns) | | $\sigma$ (ns) | |
|---|---|---|---|---|---|---|
| | Real | Fit | Real | Fit | Real | Fit |
| Fig 3 (e), Blue curve | 1.133 | 1.156 [0.075] | 800 | 752 [9] | 200 | 200 [14] |
| Fig 3 (e), Red curve | 1.889 | 1.620 [0.092] | 800 | 759 [9] | 200 | 235 [14] |
| Fig 3 (f), Green curve | 1.889 | 1.781 [0.067] | 600 | 562 [5] | 200 | 213 [9] |
| Fig 3 (f), Blue curve | 1.889 | 1.620 [0.092] | 800 | 759 [9] | 200 | 235 [14] |
| Fig 3 (f), Red curve | 1.889 | 1.772 [0.071] | 1000 | 968 [6] | 200 | 212 [9] |
| Fig 3 (g), Green curve | 1.889 | 1.620 [0.092] | 800 | 759 [9] | 200 | 235 [14] |
| Fig 3 (g), Blue curve | 1.889 | 1.846 [0.038] | 800 | 758 [4] | 300 | 304 [7] |
| Fig 3 (g), Red curve | 1.889 | 1.869 [0.027] | 800 | 779 [4] | 400 | 407 [8] |

TABLE S2. Summary of fit parameters obtained from data reported in Fig. 3 of the main text, exluding the $B_0 = 0$ mT curve, for the Gaussian-shaped extrernal field. Numbers given in square brackets correspond to the errors obtained from the fit procedure.



\end{document}